\documentclass[10pt,aps,prl,twocolumn,showkeys,superscriptaddress]{revtex4-2}

\usepackage{placeins}                   
\usepackage[utf8]{inputenc}
\usepackage{newunicodechar}
\newunicodechar{，}{,}
\usepackage[artemisia]{textgreek}       
\usepackage{amssymb}       
\usepackage{gensymb}
\usepackage{amsmath} 				    
\usepackage{latexsym}                   
\usepackage{nicefrac}                   
\usepackage{float}
\usepackage{booktabs} 
\usepackage{graphicx}                   
\usepackage{epsfig}                     
\usepackage{epstopdf}                   
\usepackage{dcolumn}                    

\usepackage{lipsum}                     
\usepackage{color}                      

\usepackage{subfigure}                  
\usepackage{blindtext}                  
\usepackage[ulem=normalem]{changes}     
\usepackage{float}                      
\usepackage{array}                      
\usepackage{soul}                       
\usepackage[T1]{fontenc}                
\usepackage[separate-uncertainty=true,multi-part-units=single]{siunitx}
\usepackage{siunitx}
\usepackage[hidelinks]{hyperref}        

\DeclareSIUnit\sq{\ensuremath{\Box}}
\DeclareSIUnit\bar{bar}
\DeclareSIUnit\angstrom{\text{Å}}
\DeclareSIUnit{\atpercent}{at\%}

\newcolumntype{P}[1]{>{\centering\arraybackslash}p{#1}}
\hypersetup{
    colorlinks,
    linkcolor={red!50!black},
    citecolor={blue!50!black},
    urlcolor={blue!80!black}
}

\newcolumntype{L}[1]{>{\raggedright\arraybackslash}p{#1}}
\newcolumntype{C}[1]{>{\centering\arraybackslash}p{#1}}
\newcolumntype{R}[1]{>{\raggedleft\arraybackslash}p{#1}}

\begin{document}

\title{Impact of Layer Structure and Strain on Morphology and Electronic Properties of InAs Quantum Wells on InP (001) }

\author{Zijin Lei}
\email{zijin.lei2026@gmail.com}

\affiliation{Solid State Physics Laboratory, ETH Z\"urich, CH-8093 Z\"urich, Switzerland}
\affiliation{Quantum Center, ETH Z\"urich, CH-8093 Z\"urich, Switzerland}

\author{Yuze Wu}
\affiliation{Solid State Physics Laboratory, ETH Z\"urich, CH-8093 Z\"urich, Switzerland}
\affiliation{Quantum Center, ETH Z\"urich, CH-8093 Z\"urich, Switzerland}

\author{Christian Reichl}
\email{creichl@phys.ethz.ch}
\affiliation{Solid State Physics Laboratory, ETH Z\"urich, CH-8093 Z\"urich, Switzerland}
\affiliation{Quantum Center, ETH Z\"urich, CH-8093 Z\"urich, Switzerland}

\author{Stefan Fält}
\affiliation{Solid State Physics Laboratory, ETH Z\"urich, CH-8093 Z\"urich, Switzerland}
\affiliation{Quantum Center, ETH Z\"urich, CH-8093 Z\"urich, Switzerland}

\author{Werner Wegscheider}
\affiliation{Solid State Physics Laboratory, ETH Z\"urich, CH-8093 Z\"urich, Switzerland}
\affiliation{Quantum Center, ETH Z\"urich, CH-8093 Z\"urich, Switzerland}

\begin{abstract}
High-quality InAs quantum wells grown on InP are a promising platform for topological quantum information processing due to their large \textit{g}-factor, strong Rashba spin–orbit interaction, and their compatibility with in-situ–deposited superconductors. In this work, we investigate InAs/InGaAs quantum wells grown on InP (001) wafers, focusing on how the layer structure and strain influence the electronic properties and surface morphology. By combining quantum transport measurements with atomic force microscopy, we show that the layer design predominantly affects the mobility anisotropy, which aligns well with the surface morphology. Surface characterization further reveals the mechanism of quantum well collapse when the layer thickness exceeds the strain limit. In addition, transport measurements demonstrate that quantum confinement has a clear impact on band nonparabolicity.  
\end{abstract}

\maketitle
As a well-known narrow-bandgap III–V semiconductor, InAs holds high electron mobility, a small effective mass, a large \textit{g}-factor, and strong Rashba spin–orbit interactions  \cite{Madelung,Winkler2003,vurgaftman2001band}. In recent years, InAs quantum wells (QWs) have attracted increasing attention because they can be proximitized to be superconductive by an in-situ deposited superconductor \cite{cheah2023control,telkamp2025development}. This makes InAs QWs a promising platform for studying topological superconductivity in integrated multi-terminal devices \cite{fornieri2019evidence,coraiola2023phase,coraiola2024spin,coraiola2024flux}.
However, growing high-quality InAs QWs is considerably more challenging than growing standard GaAs structures. A significant difficulty is the lack of a mature InAs substrate. At present, InAs QWs are typically grown on GaAs, InP, and GaSb substrates. Although GaAs is most available, its large lattice mismatch with InAs severely limits the achievable mobility \cite{kirti2025optimization}. In contrast, InAs QWs grown on GaSb, where the lattice constants are closely matched, offer the highest reported mobilities, where mobility $\mu$ reaching $2.4\times\: \rm{10^6 cm^/Vs}$ at a density of $n = 1\times 10^{12}\: \rm{cm^{-2}}$ \cite{Tschirky2017}. However, these QWs exhibit trivial edge states, which complicates the fabrication of advanced devices \cite{mittag2017passivation}. Therefore, transport studies, such as the fractional quantum Hall effect and nanoscale constrictions, require complex gate architectures \cite{ma2017observation,mittag2018edgeless,mittag2019gate,komatsu2022gate}. For comparison, InP-based InAs/InGaAs QWs have become a widely used platform for nanostructures and topological superconductor devices. The mobility of deep InAs QWs can reach above $10^6\:\rm{cm^2/Vs}$ \cite{hatke2017mobility,dempsey2025effects}. In shallow QW structures, InAs two dimensional electron gas (2DEG) can be easily proximitized by in-situ deposited Al or Al/Nb films \cite{cheah2023control,telkamp2025development}. Furthermore, recent studies on quantum point contacts have shown that InAs QW based on InP wafers do not have trivial edge states, which facilitates the research of quantum devices based on 2D InAs \cite{hsueh2022clean}.

However, the active region of these InGaAs/InAs heterostructures carries significant strain, which limits the maximum achievable QW width. As a result, interface scattering becomes a key factor restricting electron mobility. Recently, Dempsey \textit{et al}. \cite{dempsey2025effects}. introduced a refined strain-engineering approach in both the buffer and cladding layers, substantially increasing the allowable well width and thereby enhancing the peak mobility up to $10^6 \: \rm {cm^2/Vs}$. Despite these advances, the influence of strain and quantum confinement on key properties of InAs QWs, such as mobility anisotropy and band nonparabolicity, remains an important topic for further investigation.

In this work, we investigate strained InAs QWs grown on InP substrates, focusing on the effects of barrier and well thickness on surface morphology and 2DEG properties. By tuning the InGaAs cladding thickness and the QW width, we demonstrate high-mobility 2DEGs with peak mobilities up to $1.03\times 10^6\: \rm {cm^2/Vs}$ at 10 mK. AFM measurements reveal the correlation between surface morphology and transport anisotropy, and identify the structural origin of QW collapse when the layer thickness exceeds the strain limit. Furthermore, the effective mass has been extracted from temperature-dependent Shubnikov-de Haas (SdH) oscillation measurements in both low- and high-density regimes, showing strong band nonparabolicity in narrow wells. 

The QW samples in this work were grown in a modified Veeco Gen II molecular beam epitaxy (MBE) system. The growth procedure is similar to the approach reported in Ref. \cite{hatke2017mobility,dempsey2025effects} . First, Fe-doped InP (001) semi-insulating substrates were thermally desorbed in vacuum. The wafer was then loaded into the growth chamber. It is worth noting that the first epitaxial layer ($\rm{In_{0.58}Al_{0.42}As}$) was grown at 480 °C after deoxidizing the InP surface in an As environment at 510 °C. The deoxidation process must be carefully controlled to avoid the formation of InAs, which would lead to the failure of subsequent growth. After covering the surface with lattice-matched $\rm{In_{0.58}Al_{0.42}As}$, the superlattice (SL) of $\rm{In_{0.575}Al_{0.425}As}/In_{0.59}Al_{0.41}As$ was grown at a relatively lower temperature (< 390 °C). We note that precise control of the $\rm{Al:In}$ ratio in this SL is crucial for achieving a high-mobility 2DEG in the QW. Following the established procedure in previous publications \cite{cheah2023control}, step-graded buffer layers were grown, each increasing the In content by $1.35\%$ and the total thickness was up to 1 $\rm{\mu m}$ . Next, a 58-nm-thick $\rm{In_{0.77}Al_{0.23}As}$ layer was grown at 420 °C as a virtual substrate. On this virtual substrate, the cladding layer of  $\rm{In_{0.75}Ga_{0.25}As}$ was grown as the bottom barrier of QW, with the thickness $d$ set to 10.5 or 12 nm. The InAs layer with width $w$ was then grown as the QW, followed by the top $\rm{In_{0.75}Ga_{0.25}As}$ cladding layer, which is identical to the bottom barrier. Finally, a 120-nm-thick $\rm{In_{0.75}Al_{0.25}As}$ capping layer was grown to minimize surface scattering. No additional doping was introduced during the growth. In this work, five samples are grown and characterized for comparison. After growth, all samples were first characterized  at 4.2 K using the standard 5 mm × 5 mm van der Pauw (vdP) geometry with In-soldered contacts. The controlled parameters ($d$ and $w$) of each wafer, along with the vdP density $n_{\rm{vdP}}$ and mobility $\mu _{\rm{vdP}}$, are summarized in Table \ref{table:vdp}.

\begin{table}
    \centering
    \begin{tabular}{ccccc}\toprule
         Sample&  $d$ (nm)&  $w$ (nm)&  $n_{\rm{vdP}}$ ($10^{11} \rm{cm^{-2}}$)&  $\mu_{\rm{vdP}}$ ($10^{5} \rm{cm^2/Vs}$)\\\midrule
 A& 12& 2& 2&2.7\\
 B& 12& 6& 2.2&5.4\\
 C& 12& 10& 2&2.5\\
 D& 10.5& 10& -&-\\
 E& 10.5& 16& -&-\\ \bottomrule\end{tabular}
    \caption{Summary of the vdP characterization at 4.2 K}
    \label{table:vdp}
\end{table}

\begin{figure}
    \centering
    \includegraphics[width=1\linewidth]{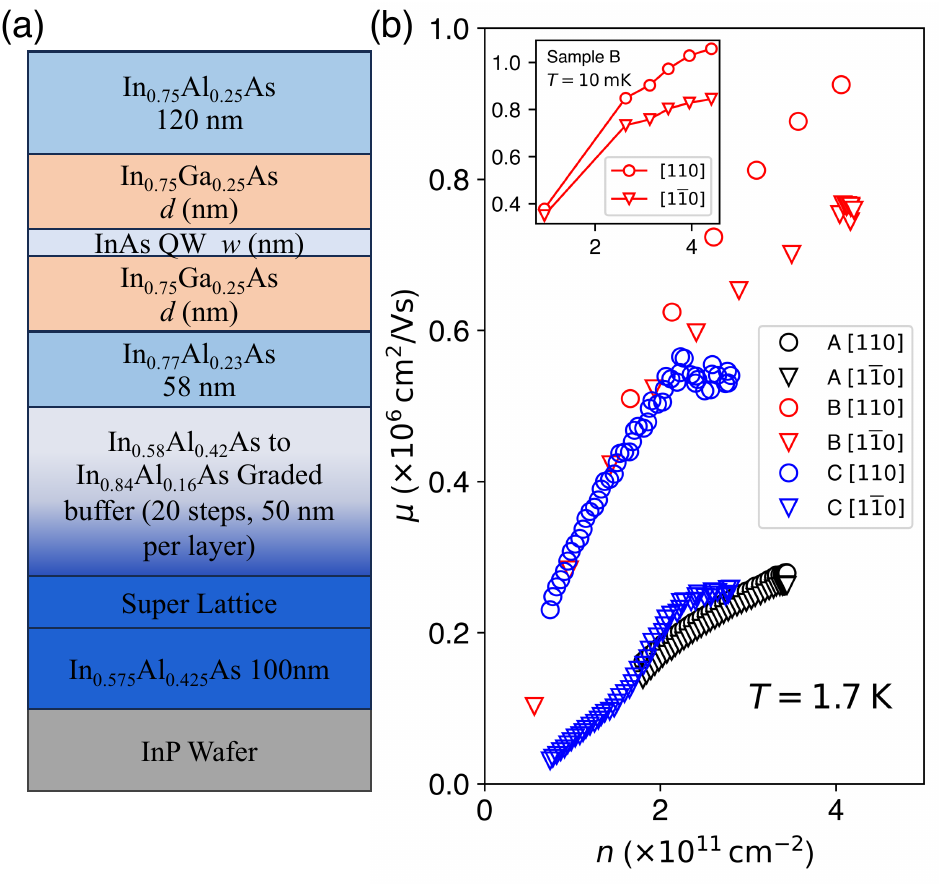}
    \caption{(a) Schematic  image of InAs QW layer-structure. (b) Electron mobility vs density of Sample A, B and C along $[110]$ and $[1\overline{1}0]$ directions, measured from "L"-shape Hall bars with global top gates at 1.7 K. Inset: Same Hall bar based on Sample B, but measured at 10 mK, showing the peak mobility reaching $1.03\times 10^6 \:\rm{cm^2/Vs}$.}
    \label{fig:mobility}
\end{figure}

As shown in Fig. \ref{fig:mobility}, the electron mobility increases sub-linearly with increasing carrier density. The dielectric layer limits the maximum achievable 2DEG density. For the same $n$, Sample B exhibits the highest mobility. At 1.7 K, the maximum value of $\mu = 0.98 \times \rm{10^6 cm^2/Vs}$ is achieved along $[1 1 0]$ with a density of $n = 4.5\times10^{11}\:\rm{cm^{-2}}$. The increase of mobility saturate at higher density regimes is due to stronger interface scattering as the wave function becomes more asymmetric. The peak mobility of Sample B continues to increase at lower temperature. At 10 mK, the peak mobility reaches $1.03\times 10^6\: \rm{cm^2/Vs}$ along the $[110]$ direction, achieving previous records by Hatke \textit{et al}. with a lower density \cite{hatke2017mobility}, and it is comparable with the QW by Dempsey \textit{et al}. where strain engineering was performed \cite{dempsey2025effects}. 

In addition to the mobility differences, Samples A, B, and C exhibit distinct anisotropies. Although mobility along $[1 1 0]$ is always higher than along $[1 \bar{1} 0]$, anisotropy increases with increasing QW thickness. A QW width of $w = 6$ nm offers the best balance, giving the highest mobility while maintaining moderate anisotropy. Because the carrier density remains low throughout the full range of $n$, no signatures of second-subband occupation were observed.

\begin{figure*}
 \centering
    \includegraphics[width=1\linewidth]{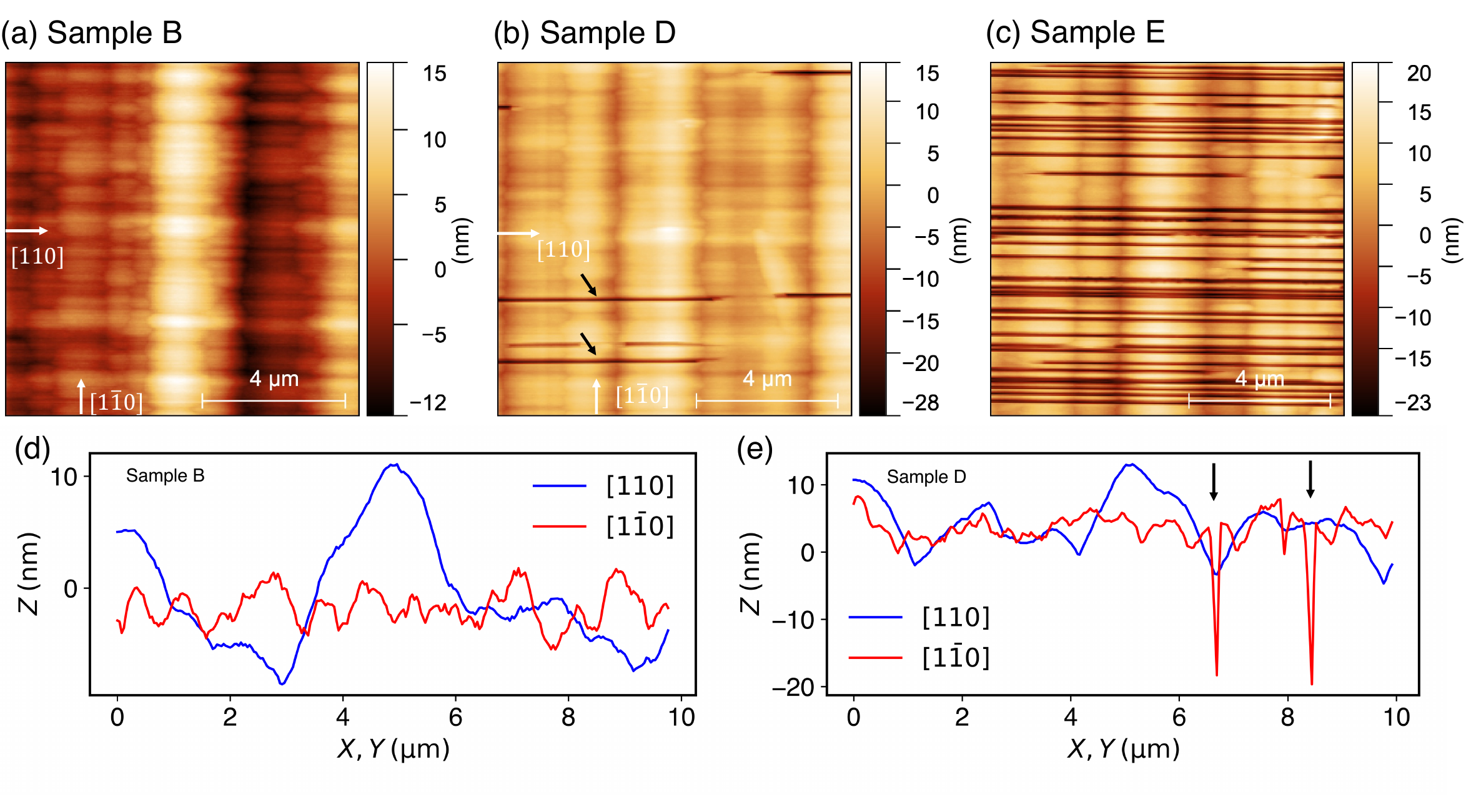}
    \caption{(a), (b) and (c) Successive AFM surface characterization of Sample B, D and E. The measurement range is $10\:\rm{\mu m}\times 10\: \rm{\mu m}$. Crystal directions $[110]$ and $[1\overline{1}0]$ are labeled. (d) and (e) are line cut (white arrows) of Sample B and D along $[110]$ and $[1\overline{1}0]$. The appearing deep grows along $[110]$ are labeled with black arrows in both (b) and (e).}
    \label{fig:AFM}    
\end{figure*}

Figures \ref{fig:AFM}(a) to (c) present the AFM surface characterization of Samples B, D, and E, respectively. In general, all samples show clear crosshatch patterns with stripes along the $[110]$ and $[1 \bar{1} 0]$ directions. Comparing the line cuts shown in $[1 1 0]$ and $[1 \bar{1} 0]$ (Figs. \ref{fig:AFM}(d) and (e)), although the peak-to-peak roughness in the $[1 1 0]$ direction is much larger and can even be comparable to the QW thickness, the correlation length is much larger. Therefore, the higher mobility along the $[1 1 0]$ direction is attributed to the large correlation length, which gives a higher limit of the mean free path of the electrons. More data are presented in the Appendix.

The surface morphology may also explain the collapse of the QW in Samples D and E. Although the thickness of $\rm{In_{0.75}Ga_{0.25}As}$ changes only by 1.5 nm, the surface of Sample D exhibits deep grooves along the $[1 1 0]$ direction, causing the QW to break into discontinuous islands. These grooves are even more obvious in Sample E, where the InAs thickness is $w=16$ nm, which is far beyond the strain limit. In addition to a large increase in groove density, the cross hatch structure is less ordered. It is worth noting that these grooves are not due to measurements, because Samples A to E were characterized successively, and the location of the grooves are the same in both large and small area measurements. See more AFM figures in the Supplementary materials. This observation is consistent with, and may explain, reports in previous work by Dempsey \textit{et al}., where the QW failed to host a 2DEG when the InAs layer was too thick \cite{dempsey2025effects}.

\begin{figure}
    \centering
    \includegraphics[width=1\linewidth]{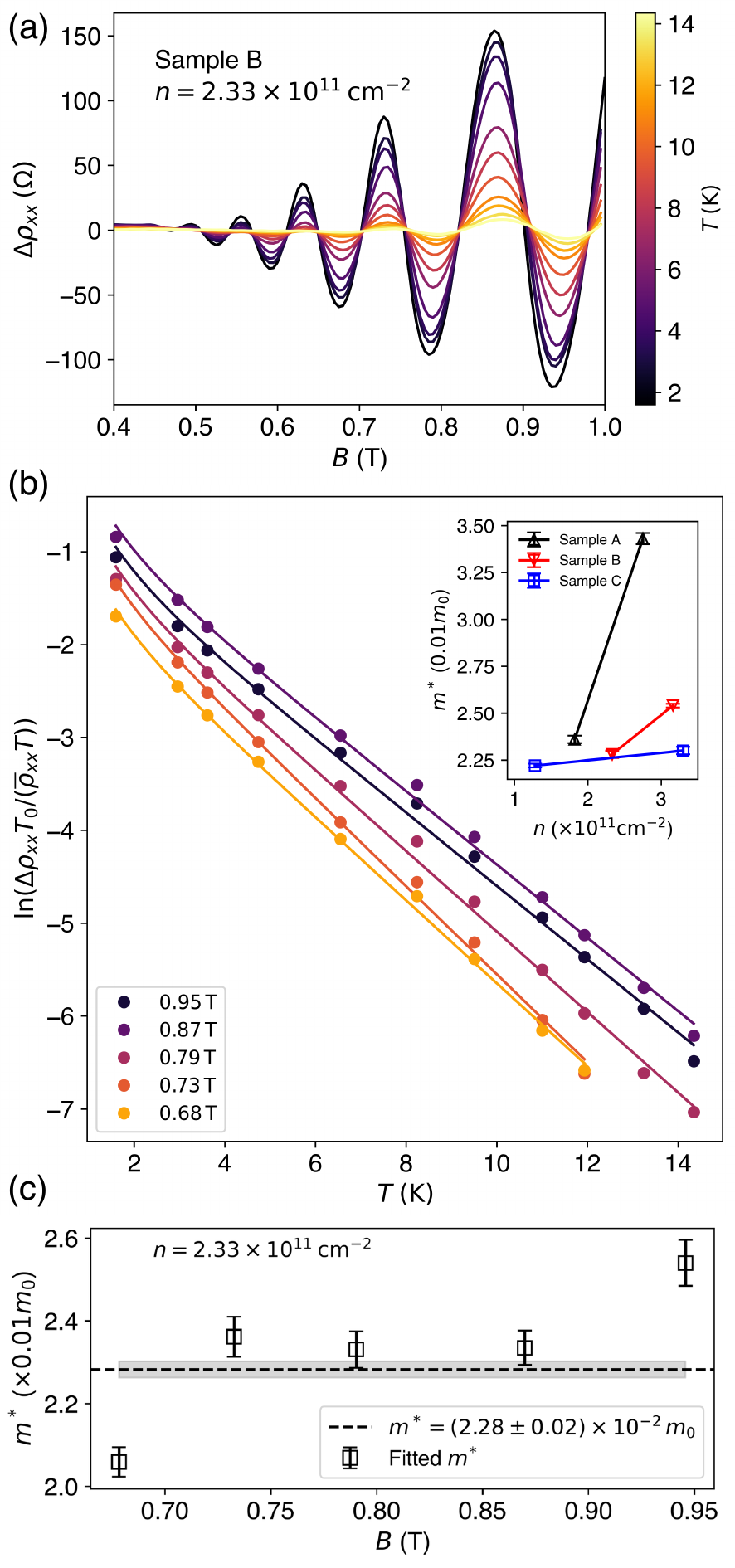}
    \caption{(a) Temperature dependence of the SdH oscillations measured on Sample B, where $n = 2.33 \times 10^{11} \rm{cm^{-2}} $. (b) Fitting of the Dingle factor based on the local maxima and minima in (a). (c) Fitted effective mass $m^*$ vs $B$. In a wide range of $B$, $m^*$ shows an average value of $(2.28\pm0.02)\times 10^{-2} m_0$ (dashed line), and the error is indicated with shadow.} 
    \label{fig:SdH}
\end{figure}

The high electron mobilities in Samples A, B, and C allow further investigation of the effect of confinement on the band structure, which can be probed through effective mass measurements. Figure \ref{fig:SdH} shows the temperature dependence of the SdH oscillations in Sample B at a carrier density of $n = 2.33\times 10^{11}\: \rm{cm^{-2}}$. The oscillatory part of the longitudinal resistivity, $\Delta \rho_{xx}$, is obtained by subtracting a polynomial background from the trace of $\rho_{xx} (B)$. Following the method introduce in Ref. \cite{lei2023gate}, we select the local maxima and minima in the $\Delta \rho_{xx}$ vs $B$ traces as data points to construct the SdH envelope. Figure 4(b) shows the fit of the Dingle factor to $\ln{(\Delta\rho_{xx} T_{0} / \overline{\rho}_{xx }T)}$, where $T_0$ is the lowest temperature at which SdH oscillations were measured, and $ \overline{\rho}_{xx }$ is the polynomial background. As shown in Fig. \ref{fig:SdH}, within a wide range of the magnetic field, the obtained effective mass is constant. Therefore, we conclude that the effective mass is $m^* =(2.28\pm 0.02)\times 10^{-2} m_0$, where $m_0$ is the electron mass in vacuum. 

Using the same methods, we extracted the effective masses of Samples A, B, and C in both the low- and high-density regimes. As shown in the inset of Fig.~\ref{fig:SdH}(b), all InAs QWs exhibit similarly small electron effective masses, close to the bulk InAs electron effective mass ($0.023m_0$). The influence of QW confinement is not obvious, in contrast to what is typically observed in standard GaAs and GaInAs QWs \cite{WetzelPhysRevB.45.14052,EkenbergPhysRevB.36.6152,HaiPhysRevB.42.11063}. Instead, reducing the QW width mainly enhances the band nonparabolicity, as evidenced by the increase of the effective mass at higher carrier densities. This effect is more pronounced in narrower QWs. This behavior is consistent with previous reports on GaAs QWs \cite{AltschulJAP,staedeleJAP}. Furthermore, strain in the QWs may also contribute to the observed behavior, as strain is known to modulate the band structure and nonparabolicity, which can affect the electron effective mass  \cite{kulbachinskii2012electron,dalfors1997effective,gauer1994energy}.

\begin{figure}
    \centering
    \includegraphics[width=1\linewidth]{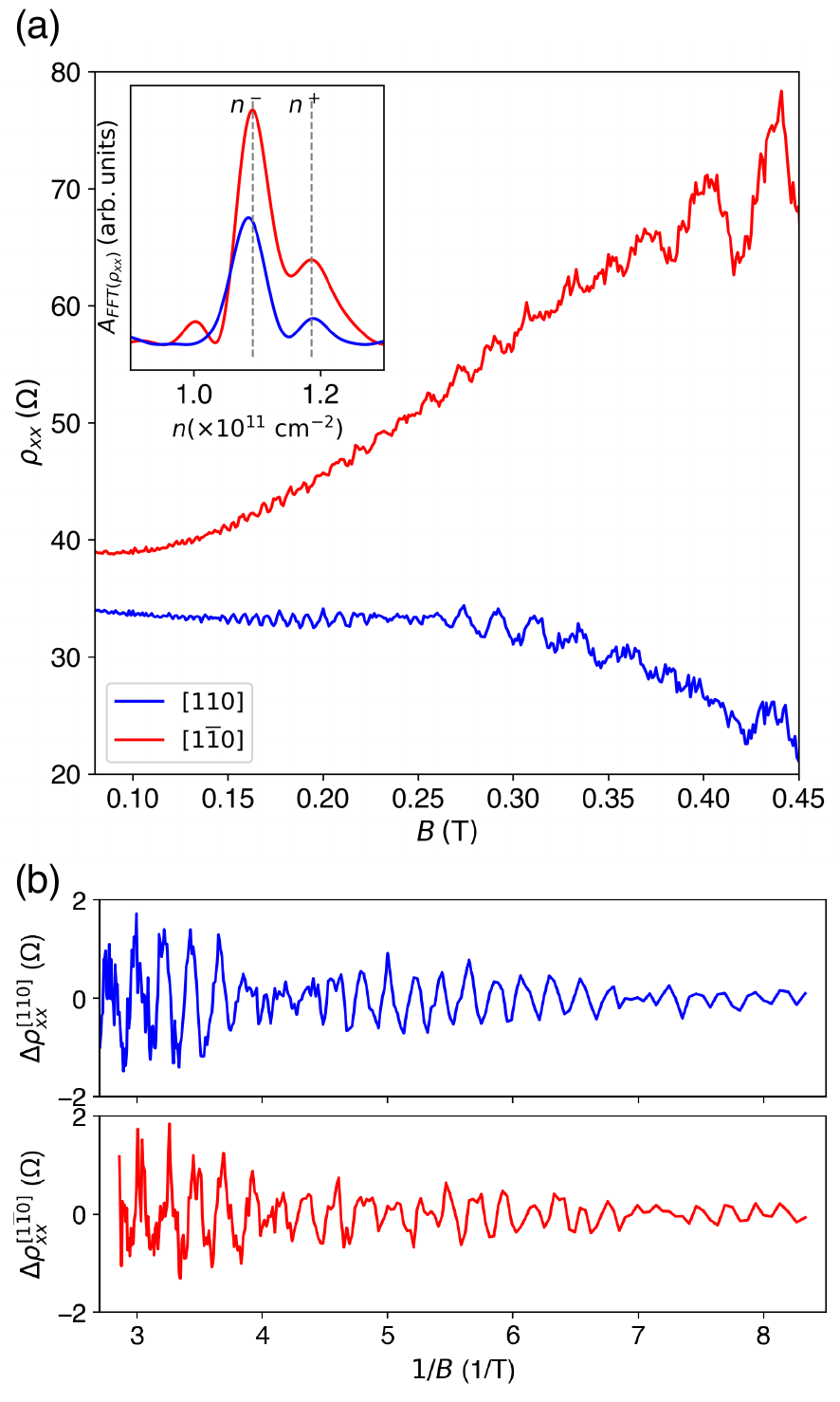}
    \caption{(a) $\rho_{xx}$ vs $B$ on both $[110]$ and$[1\overline{1}0]$ directions. Beating of the SdH oscillations are observed in both traces. \textit{inset}: the fast Fourier transform of the SdH oscillation, measured along two directions. (b) The oscillation part of the longitudinal resistivity, $\Delta\rho_{xx}$ as the function of $1/B$. Both directions are presented. } 
    \label{fig:beating}
\end{figure}

The high-mobility 2DEG in Sample B allows us to probe the spin-orbit interaction through precise SdH oscillation measurements at 10 mK. Figure \ref{fig:beating}(a) presents the SdH oscillation of Sample B with carrier density of $n = 2.1\times 10^{11}\:\rm{cm^{-2}}$ . We observed a beating of the SdH oscillations in the range between 0.25 and 0.4 T in the measurement for both $[110]$ and $[1\overline{1}0]$ directions. The oscillation parts of the longitudinal resistivity $\Delta \rho_{xx}$ of these 2 directions are shown in Fig. \ref{fig:beating}(b). With the fast Fourier transform, two peaks are visible in the power spectrum in both directions (Fig. \ref{fig:beating}) \textit{inset}. 
Here, the frequency spectrum was converted to the carrier density using the Onsager relation $n = |e|f/h$.
The corresponding carrier densities of these two peaks for the direction $[110]$ ($[1\overline{1}0]$) are $n^+ = 1.061 (1.189)\times 10^{11}\:\rm{cm^{-2}}$ and $n^- = 0.978 (1.087)\times 10^{11}\:\rm{cm^{-2}}$. With the obtained $n^+$ and $n^-$, the Rashba spin obit coefficient $\alpha _{\rm{SO}}$ can be calculated using the relation $\alpha_{\rm{SO}}=\frac{\Delta n \hbar^{2}}{m^*}\sqrt{\frac{\pi}{2(n^++n^--\Delta n)}}$, where $\Delta n = n^+-n^-$ \cite{hatke2017mobility,dempsey2025effects}. Using the effective mass obtained in Fig. \ref{fig:SdH}, the Rashba spin orbit coefficient is calculated to be 64 and 74 $\rm{meVnm}$ for $[110]$ and $[1\overline{1}0]$ directions, respectively. The agreement of the Rashba coefficients obtained from two directions strongly indicates that this beating is due to the strong SOI, not to the inhomogeneity during the epitaxy growth. Furthermore, these values agree well with previous publications \cite{hatke2017mobility,dempsey2025effects}. In addition, the beating was observed with different carrier densities. However, within the measurement precision, $\alpha_{\rm{SO}}$ did not show obvious dependence of the gate voltage. 

In summary, we studied strained InAs QWs on InP and identified how the InGaAs cladding thickness and QW width determine the structural and transport properties. AFM measurements on the surface morphology link mobility anisotropy to crosshatch-induced roughness and explain the QW breakdown in samples exceeding the strain limit. SdH oscillation measurements reveal the strong band nonparabolicity induced by confinement in the QWs. These works establish the optimal structure design for high-quality InAs QWs on InP and provide quantitative insights into the interplay between confinement, strain, and QW transport properties.

The authors thank Prof. T. Ihn for allowing us to use the low temperature setup. We acknowledge financial support from the Swiss National Science Foundation (SNSF) and the NCCR QSIT (National Center of Competence in Research - Quantum Science and Technology).

\appendix

\section{AFM Surface Characterization}

\begin{figure}
    \centering
    \includegraphics[width=0.75\linewidth]{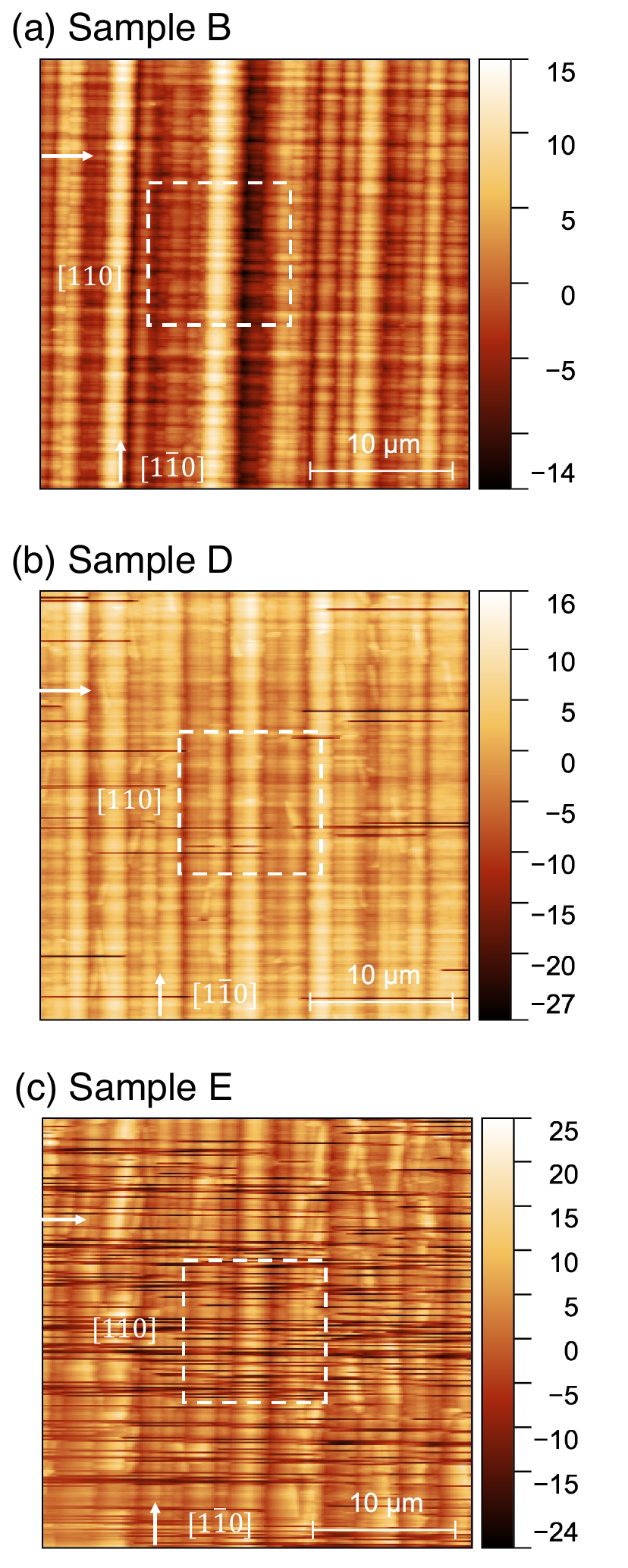}
    \caption{(a), (b) and (c) Successive AFM surface characterization of Sample B, D and E in a large scale ($\rm{30\mu m \times 30\mu m}$). Figures presented in the main contents are zoom-in measurement of the area indicated with white boxes. Crystal directions $[110]$ and $[1\overline{1}0]$ are labeled. }
    \label{fig:AFM_Large}
\end{figure}

Figure~\ref{fig:AFM_Large} presents the surface characterization of Samples B, D, and E over a larger scan area. All measurements were performed successively, and Figs.~\ref{fig:AFM}(a)–(c) are zoomed-in scans in the regime indicated with white boxes. The consistent features between the large- and small-area images confirm that the deep grooves along the $[110]$ direction are intrinsic to the sample surface. Moreover, the root mean square (RMS) roughness and the correlation lengths extracted from the central line cuts along the two crystallographic directions, $\xi_{[110]}$ and $\xi_{[1\overline{1}0]}$, are summarized in Table~\ref{tab:AFM_statistic}.

\begin{table}
    \centering
    \begin{tabular}{cccc}\toprule
     Sample    & RMS (nm) & $\xi_{[110]}$ (nm)& $\xi_{[1\overline{1}0]}$ (nm)\\\midrule
       A  & 4.51 &562  & 74\\
        B & 4.38 & 747 & 94\\
        C & 5.14 & 606 & 98\\
        D & 4.45 & 580 & 84\\
        E & 7.02 & $\sim 3000$ & 105 \\
        \bottomrule
    \end{tabular}
    \caption{Summary of the statistic values of the AFM characterization for Sample A to E}
    \label{tab:AFM_statistic}
\end{table}

\section{Effective Mass of Sample A in the High-Density Regime}

\begin{figure}
    \centering
    \includegraphics[width=1\linewidth]{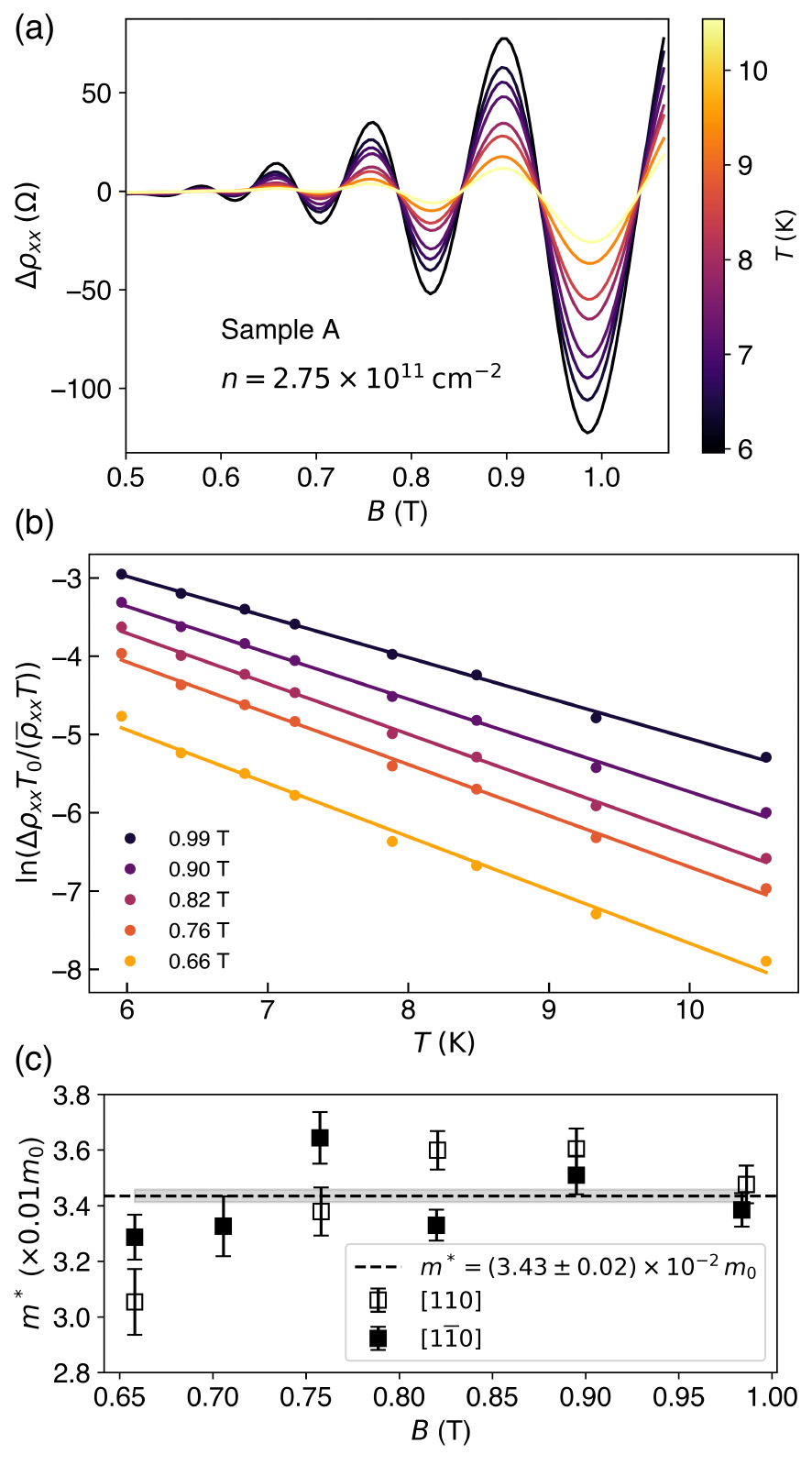}
    \caption{a) Temperature dependence of the SdH oscillations measured on Sample A along $[110]$ direction, where $n = 2.75 \times 10^{11} \rm{cm^{-2}} $. (b) Fitting of the Dingle factor based on the local maxima and minima in (a). (c) Fitted effective mass $m^*$ vs $B$. To increase the precision of the analysis, data obtained from $[1\overline{1}0]$ are also presented. the calculated effective mass is $(3.43\pm0.02)\times 10^{-2} m_0$ (dashed line).}
    \label{fig:SampleA_Mass}
\end{figure}

We present here the measurement of the effective mass of Sample A in the high-density regime, where the largest effective mass is observed. As shown in Fig.~\ref{fig:SampleA_Mass}, the temperature dependence of the SdH oscillations was measured and analyzed using the same methods described in the main text. To reduce the experimental uncertainty, data from both the $[110]$ and $[1\overline{1}0]$ directions were included. As Fig.~\ref{fig:SampleA_Mass}(c) shows, the obtained effective mass is $(3.43 \pm 0.02) \times 10^{-2} m_0$.

\bibliographystyle{ieeetr}
\bibliography{papers}

\end{document}